\documentclass{article}

\usepackage{arxiv}

\usepackage[utf8]{inputenc} % allow utf-8 input
\usepackage[T1]{fontenc}    % use 8-bit T1 fonts
\usepackage{hyperref}       % hyperlinks
\usepackage{url}            % simple URL typesetting
\usepackage{booktabs}       % professional-quality tables
\usepackage{amsfonts}       % blackboard math symbols
\usepackage{nicefrac}       % compact symbols for 1/2, etc.
\usepackage{microtype}      % microtypography
\usepackage{lipsum}		% Can be removed after putting your text content
\usepackage{graphicx}
\usepackage{natbib}
\usepackage{doi}

\title{Understanding the Influence of Receptive Field and Network Complexity in Neural-Network-Guided TEM Image Analysis}

\author{{Katherine Sytwu} \\
	Molecular Foundry\\
	Lawrence Berkeley National Laboratory\\
	Berkeley, CA 94720 \\
	\texttt{ksytwu@lbl.gov} \\
	\And
	{Catherine Groschner} \\
	Department of Materials Science and Engineering\\
	University of California Berkeley\\
	Berkeley, CA 94720 \\
	\texttt{} \\
	\And
	{Mary C. Scott} \\
	Department of Materials Science and Engineering\\
	University of California Berkeley\\
	Berkeley, CA 94720 \\
	\texttt{} \\
	Molecular Foundry\\
	Lawrence Berkeley National Laboratory\\
	Berkeley, CA 94720 \\
	\texttt{MCScott@lbl.gov} \\
}

\date{}

\renewcommand{\shorttitle}{Influence of Receptive Field}

\begin{document}

\maketitle

\begin{abstract}

Trained neural networks are promising tools to analyze the ever-increasing amount of scientific image data, but it is unclear how to best customize these networks for the unique features in transmission electron micrographs. Here, we systematically examine how neural network architecture choices affect how neural networks segment, or pixel-wise separate, crystalline nanoparticles from amorphous background in transmission electron microscopy (TEM) images. We focus on decoupling the influence of receptive field, or the area of the input image that contributes to the output decision, from network complexity, which dictates the number of trainable parameters. We find that for low-resolution TEM images which rely on amplitude contrast to distinguish nanoparticles from background, the receptive field does not significantly influence segmentation performance. On the other hand, for high-resolution TEM images which rely on a combination of amplitude and phase contrast changes to identify nanoparticles, receptive field is a key parameter for increased performance, especially in images with minimal amplitude contrast. Our results provide insight and guidance as to how to adapt neural networks for applications with TEM datasets. 

\end{abstract}
% \pacs{PACS Numbers}
% \keywords{Keywords go here}

\section*{Introduction}

Machine learning and computer vision algorithms are promising techniques to quantify and analyze the ever-increasing amount of scientific image data. Trained neural networks, in particular, have consistently outperformed traditional image analysis methods at identifying nanoparticles \citep{groschner2021machine, yildirim2021bayesian}, identifying clean graphene areas \citep{sadre2021deep}, denoising \citep{vincent2021developing}, and classifying crystal structures \citep{aguiar2019decoding} with transmission electron microscopy (TEM) data. Neural networks can, in part, associate their high performance with their ability to take on any functional form. These effective functions are influenced by both the chosen neural network architecture and the training data that dictate the learned features. 

For scientific data, it is unclear how to best customize these powerful tools as the architectures that work well for digital images of the natural world (natural images) may not be ideal for TEM images. Natural images and TEM images are inherently different in various image characteristics, including number of channels, feature sizes, and physical constraints. Additionally, large labeled datasets of natural images have been a key factor to the success of modern neural networks, allowing the traditional computer vision community to train large networks which can capture more complex behavior and deliver higher performance \citep{sun2017revisiting}. Scientific data streams, on the other hand, are often either much smaller or more expensive to label; therefore, ideal networks for scientific data are lightweight, or have relatively few trainable parameters, and consequently require less data \citep{akers2021rapid}. 

Lightweight neural networks often have smaller receptive fields, which may impact their performance. The receptive field of a network is the theoretical maximal area of the input image that the network can use to make its final decision. The receptive field is affected by the number, order, and types of layers in a neural network as well as their hyperparameters like filter and stride size, and can be calculated by: 
\begin{equation}
    RF = \sum_{\ell =1}^{L}\bigg[ (k_\ell -1)\prod_{j=1}^{\ell-1}s_j \bigg] +1
    \label{eq1}
\end{equation}
where $L$ is the number of layers, $k_\ell$ is the filter size of the $\ell$th layer, and $s_j$ is the stride of the $j$th layer \citep{araujo2019computing}. 

Recent literature has shown that receptive field is important in improving network performance when extending to new datasets, particularly those different from natural images. Modifying the receptive field to account for dataset-specific feature sizes has lead to increased performance in acoustic scene classification \citep{koutini2019receptive}, ultrasound image segmentation \citep{behboodi2020receptive}, and high-resolution TEM image denoising \citep{vincent2021developing}. Specifically with TEM images, it has been suggested that the receptive field needs to account for the larger length scales of the features of interest \citep{horwath2020understanding}, and by increasing the receptive field accordingly, researchers were able to achieve much better denoising performance \citep{vincent2021developing}. However, given the ``black-box'' nature of neural networks, it is unclear how to best utilize this knowledge without extensive trial and error. 

In this paper, we systematically explore how neural network architecture choices affect image analysis performance in both low-resolution and high-resolution TEM images. By varying the receptive field without changing the number of network parameters, we decouple the effects of network complexity, receptive field, and training dataset. Specifically, we focus on the task of pixel-wise separating crystalline nanoparticles from an amorphous background, a task also known as image segmentation. TEM image segmentation is a highly time-consuming task that has been difficult to automate with traditional image processing techniques like thresholding and is a useful first step for various other analysis pipelines, including calculating nanoparticle size distributions \citep{yildirim2021bayesian}, classifying defects \citep{groschner2021machine}, or tracking nanoparticles over time in \textit{in situ} datasets \citep{yao2020machine}. Segmentation is also an ideal task to further unravel network behavior as it naturally identifies image regions that the neural network struggles to understand. Our work provides intuition behind network architecture choices, hopefully moving hyperparameter decisions away from a trial-and-error process to a more informed process. 

\section*{Materials and Methods}
\subsection*{Dataset Aquisition}

2.2nm Au nanoparticles with citrate ligands were purchased from Nanopartz. 5nm, 10nm, and 20nm Au nanoparticles capped with tannic acid were purchased from TedPella. To create the TEM sample, 5uL of the nanoparticle solution was dropcasted onto ultrathin carbon TEM grids from TedPella, let sit for about 5 minutes, and then excess liquid was wicked off with a Kimwipe. 

High-resolution TEM images of the 2.2nm, 5nm, and 10nm Au nanoparticles were acquired using an aberration-corrected TEAM 0.5 TEM at 300kV. High-resolution images were 4096x4096 pixels in size at an approximate dosage of 423e/\AA$^2$. Low-resolution TEM images of 20nm Au nanoparticles were taken with a non-aberration corrected TitanX TEM at 300kV. Low-resolution images were 2048x2048 pixels in size at an approximate dosage of 16e/\AA$^2$. 

\subsection*{Dataset Creation}
Each image was manually segmented and labeled using LabelBox. For preprocessing, pixel outliers from x-rays were detected and removed, and then each image standardized (mean set to 0 and standard deviation set to 1). Images were then split up into 512x512 pixel patches to reduce memory requirements during training. Patches that only consisted of amorphous background were removed from the dataset to avoid class imbalance issues during training. Dataset characteristics are summarized in Table \ref{tab:dataset_table}, with dataset size referring to the number of unique patches. 
\begin{table}[]
    \centering
    \begin{tabular}{|c|c|c|c|c|}
    \hline
    Used in Figure(s) & Nanoparticle Diameter & Pixel Size & Dataset Size & Source\\
    \hline
    2,4,6,7,8 & 5 nm & 0.02152 nm & 216 & Groschner et al. (2021)\\
    \hline
    2,5 & 20 nm & 0.1243 nm & 132 & This paper \\
    \hline
    3 & 2.2 nm & 0.02 nm & 355 & This paper \\ 
    \hline
    3 & 5 nm & 0.02 nm & 211 & This paper\\
    \hline
    3 & 10 nm & 0.02 nm & 128 & This paper\\
    \hline
\end{tabular}
    \caption{Datasets used in this paper. Dataset size refers to the number of unique 512x512 pixel patches before augmentation.}
    \label{tab:dataset_table}
\end{table}

The patches were then split 70-10-20 into training, validation, and test sets, ordered such that patches from the same image were not likely to be in both the training and test sets. Each set was then augmented with the 8 dihedral transformations, and then randomly shuffled. 

\subsection*{Computational Framework}

\begin{figure}
    \centering
    \includegraphics[width=\textwidth]{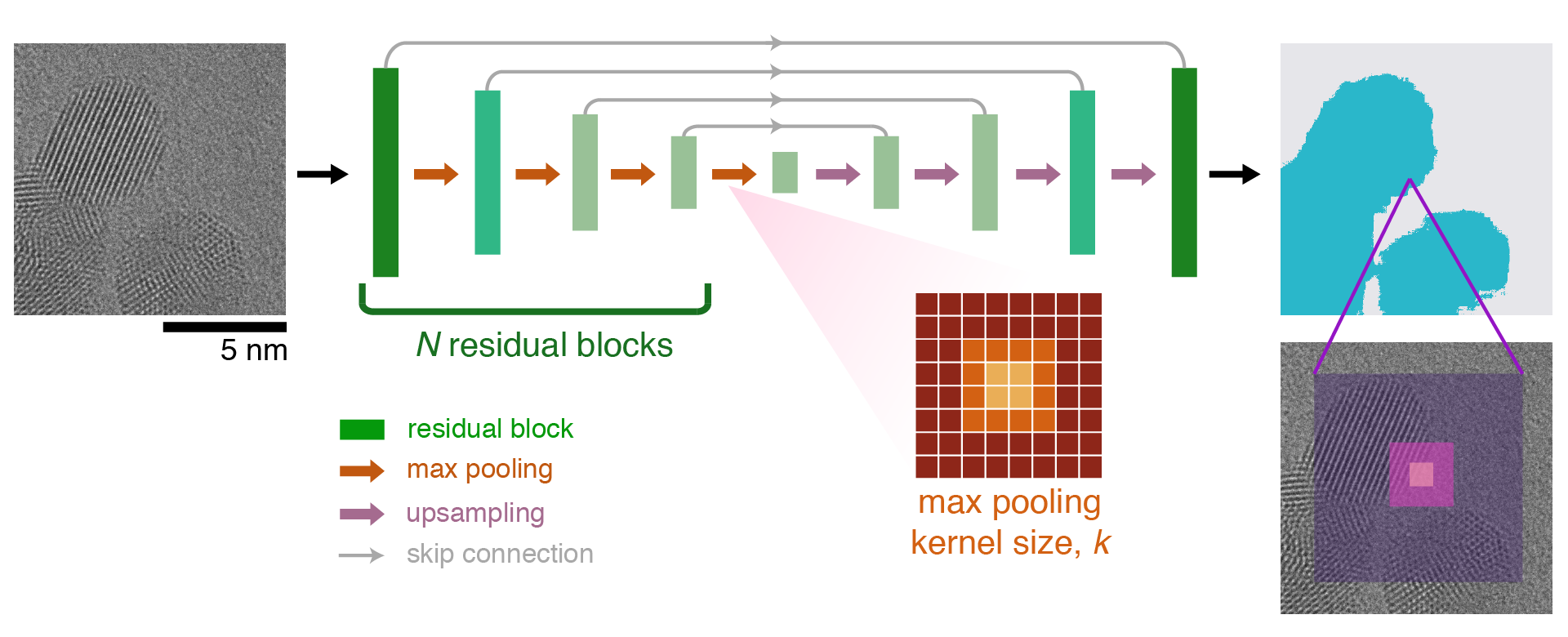}
    \caption{Overview of the UNet-based neural network architectures. We track segmentation performance as we vary the receptive field by either changing the number of residual blocks, $N = 2,3,4$, and/or the max pooling kernel size, $k = 2,4,8$. }
    \label{fig:Unet}
\end{figure}

Our network architecture is a UNet structure constructed with residual blocks. The UNet architecture is commonly used in image segmentation, consisting of a contracting encoder arm and an expansive decoder arm, with the two arms mirroring each other in structure and connected to one another via skip connections (Figure 1) \citep{ronneberger2015u}. Each arm is composed of $N$ residual blocks connected by max pooling layers which downsample the residual block output features. The residual block structure consists of a convolutional layer, followed by a batch norm layer, then a rectified linear unit (ReLU) layer, and then repeated. A skip connection connects the input to the final ReLU layer, creating a structure where the first five layers are learning the "residual" between the input and the output \citep{he2016deep}. We set the number of filters in each convolutional layer to be constant for each residual block, starting with 4 filters and doubling with each new residual block. The size of each convolutional filter is kept constant at 3x3 pixels with a stride of 1 pixel. 

To change the receptive field, we vary the max pooling filter size (and the corresponding upsampling filter size) to be either $k=$2, 4, or 8 pixels. Modifying the max pooling filter size only affects the receptive field since the max pooling layer does not have any trainable parameters. Therefore, we can construct networks that share the same number of parameters (i.e. complexity) but have varying receptive fields. As the stride ($s$) of the max pooling layer scales with the filter size, this leads to a much wider range of receptive fields (see Equation \ref{eq1}) than simply modifying filter size. We restrict our study to architectures with receptive fields smaller than the total image size, and have provided the calculated receptive fields for all network architectures in Table S1. In this paper, we present the receptive field size in nanometers rather than pixels to better contextualize the receptive field size in relation to nanoparticle features.  

\begin{table}[]
    \centering
    \begin{tabular}{|c|c|}
    \hline
        \# of Residual Blocks & \# of Trainable Parameters \\
        \hline
         $N = $ 2 & 8074\\
         \hline
         3 & 32730 \\
         \hline
         4 & 130682 \\
         \hline
    \end{tabular}
    \caption{The three architectures used in this paper, and their number of trainable parameters.}
    \label{tab:network_complex}
\end{table}

To understand how network complexity affects our results, we utilize three architectures with different numbers of residual blocks from $N$=2 to 4, with the total number of trainable network parameters shown in Table \ref{tab:network_complex}. More complex, or deeper, neural networks perform better as deeper architectures can better construct functions that capture non-linear image features. By comparing a lightweight network ($N=2$) against a more traditional deep UNet structure ($N=4$), we can identify to what extent more complexity is needed.

\subsection*{Network Training and Evaluation}
When comparing across different neural networks, the augmented training set, the order in which the network sees batches of images, and the initialized weights are all kept constant such that the only differences between networks are the architectural choices. The validation set is used to determine the number of training epochs for each dataset to prevent overfitting; networks were trained for either 100 or 150 epochs. 

Networks are trained with a cross-entropy loss function which pixel-wise penalizes the network for predictions far from ground truth, and with the Adam optimizer using a learning rate of $10^{-4}$. Each network is trained 5 times with different initialized weights, and the reported performance is the average and standard deviation of those 5 runs on the test set. Training was done either locally on a Nvidia RTX3090 GPU or on a cluster with a Nvidia K80 GPU.

Segmentation performance is evaluated by the dice score which measures the similarity between two images and quantifies it between 0 and 1, with 1 being a perfect replication of the ground truth label. We are primarily interested in the networks' ability to identify nanoparticles, and so we treat this as a binary prediction, and calculate the dice score as follows:
\begin{equation}
    D = \frac{2|X \cap Y|}{|X| + |Y|} 
\end{equation}
% = \frac{2TP}{2TP+FP+FN}
where $X$ is the predicted segmentation, $Y$ is the ground truth, and the $| \: |$ operation calculates the number of pixels classified to be a nanoparticle.  Since there are only two classes (nanoparticle and background), the dice score penalizes undersegmentation or false negatives (missing an area that is labeled as nanoparticle) more than oversegmentation or false positives (classifying background as nanoparticle) (see SI for proof). This makes the dice score a useful metric for nanoparticle segmentation because missing nanoparticle regions is a more dire consequence as false positives can be eliminated later on in an image analysis pipeline. The dice score can be calculated in two ways: either using the binary predictions (hard dice score) or using the predicted probabilities of each class (soft dice score). In this paper, we use the hard dice score to measure performance, but also report the soft dice scores in the supplementary information, which give a better indication of how confident a network is in its prediction. 

\subsection*{Fourier Filtering}
To Fourier filter the high-resolution TEM images, we apply a bandpass filter to the fast Fourier transform (FFT) of the images. The bandpass location and width are chosen such that they capture the dominant 1st order Fourier peaks in the FFT. The masked FFT is then inverted, and then blurred with a Gaussian filter (9 pixel filter size). For each image, the threshold value is determined by Otsu's method, and all pixels above the threshold value are classified as nanoparticle.  

\subsection*{Dilated Convolution}
Another strategy to increase receptive field is to dilate the convolution filters which increases the filter size without changing the number of filter pixels. Dilation is quantified by a parameter $\alpha$ which sets the spacing between pixels within the convolution filter. As noted by \cite{araujo2019computing}, to calculate the receptive field with dilated convolution layers, one just replaces the filter size $k_\ell$ in Equation \ref{eq1} with $\alpha(k_\ell -1) +1$. We set the dilation parameter to be constant within each residual block, and the exact architectural parameters are given in Table S2. 

\section*{Results}

From an image analysis perspective, there are two regimes of TEM imaging: low-resolution and high-resolution. In low-resolution TEM images, nanoparticles are primarily identified using image contrast as a result of amplitude contrast; nanoparticles appear dark against a bright background (Figure 2a). On the other hand, in high-resolution TEM images, nanoparticle amplitude and phase contrast lead to slightly darker regions with visible lattice fringes; nanoparticles are then distinguished from the background using both image contrast and image texture (Figure 2b). Low-resolution TEM images, then, require a network that can detect changes in image contrast, while high-resolution TEM images need a network that can both detect macroscopic changes in image contrast and distinguish between amorphous and crystalline textures. Therefore, networks trained for these distinct image tasks will likely behave differently and have distinct characteristics. 

\begin{figure}
    \centering
    \includegraphics[width=0.8\textwidth]{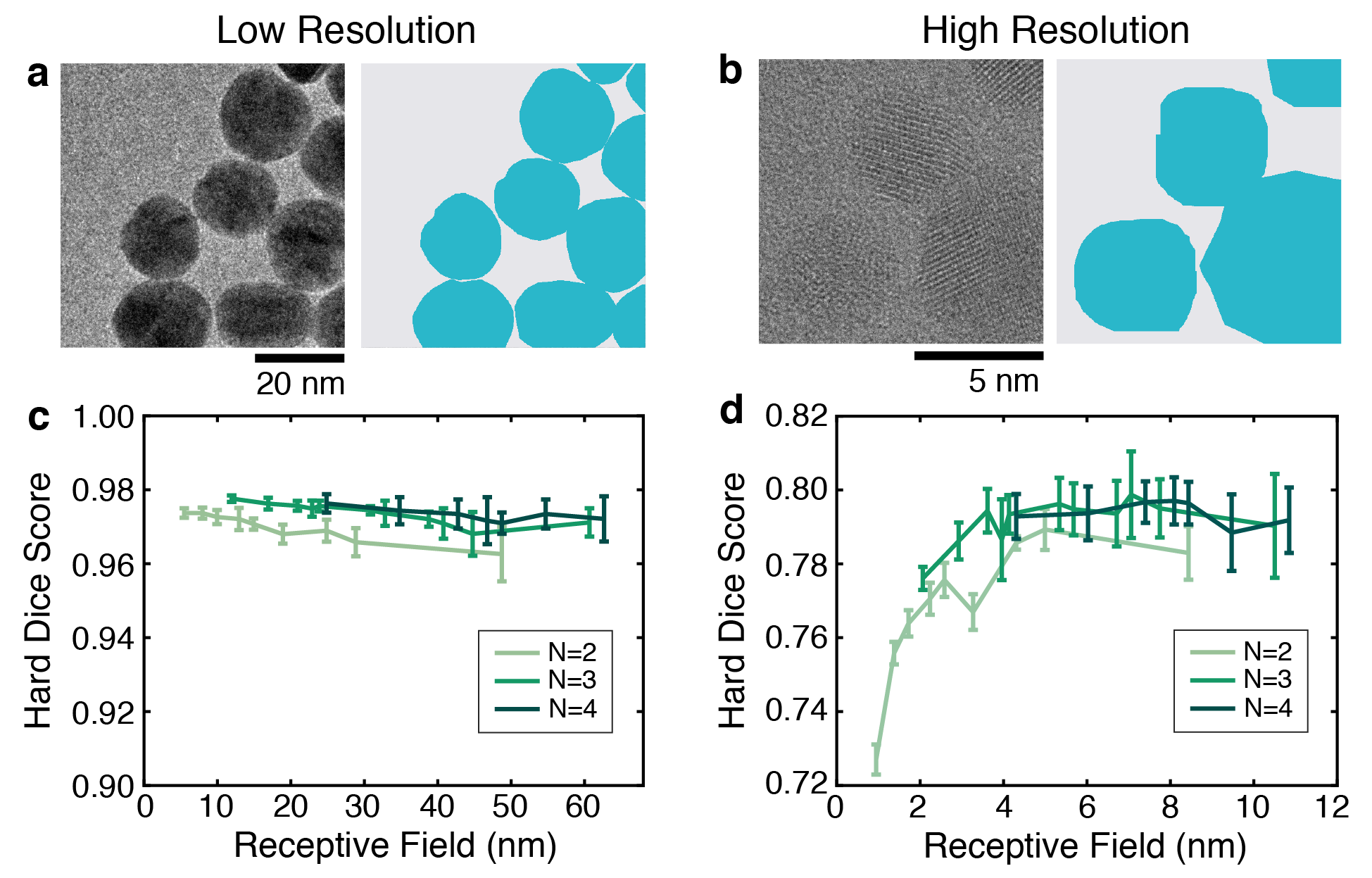}
    \caption{Receptive Field dependence on low-resolution and high-resolution TEM images. (a,b) Example images and ground truth labels from the (a) low-resolution TEM image dataset of 20nm Au nanoparticles and (b) high-resolution TEM image dataset of 5nm Au nanoparticles. (c,d) Segmentation performance as receptive field is increased for the (c) low-resolution dataset and (d) high-resolution dataset. Results are plotted for three different network complexities. }
    \label{fig2}
\end{figure}

We first examine how receptive field affects networks that are trained to segment low-resolution TEM images. We train three neural network architectures with different complexities ($N=2, 3, $ or 4) on a 20nm Au nanoparticle dataset and vary the receptive field for each architecture. As the receptive field is increased, segmentation performance remains high, with only a slight decrease in performance at large receptive fields (Figure 2c). For a simple 2-residual-block network, the dice score is $0.974\pm 0.001$ when the receptive field is 5.5nm or a quarter of the nanoparticle diameter, and then becomes $0.963\pm 0.007$ when the receptive field is 48.7nm, or over twice the average nanoparticle size. Increasing complexity leads to a slight increase in performance, with the 3-residual block and 4-residual block networks outperforming the 2-residual block network at all receptive field values.  

On the other hand, when we repeat the same training but with high-resolution TEM images of 5nm nanoparticles, we notice a stronger dependence on receptive field. For this contrast and texture-based segmentation task, performance increases with larger receptive fields but then plateaus at a certain receptive field size. Again, for a 2-residual-block network, the dice score starts at $0.727\pm0.004$ for a small receptive field of 0.95nm, but then increases to $0.783\pm0.007$ for a large receptive field of 8.4nm. This plateauing trend is seen in both the 2-residual block and 3-residual block networks, but not for the 4-residual block networks as it starts off with a receptive field around the plateau region. In contrast to the low-resolution TEM images, both complexity and receptive field influence segmentation performance in high-resolution TEM images. Given the same receptive field, a more complex network may perform better. However, a simpler network with a large enough receptive field can outperform a more complex network with a smaller receptive field. This suggests that receptive field is an important consideration when working with high-resolution TEM images. 

\begin{figure}
    \centering
    \includegraphics[width=\textwidth]{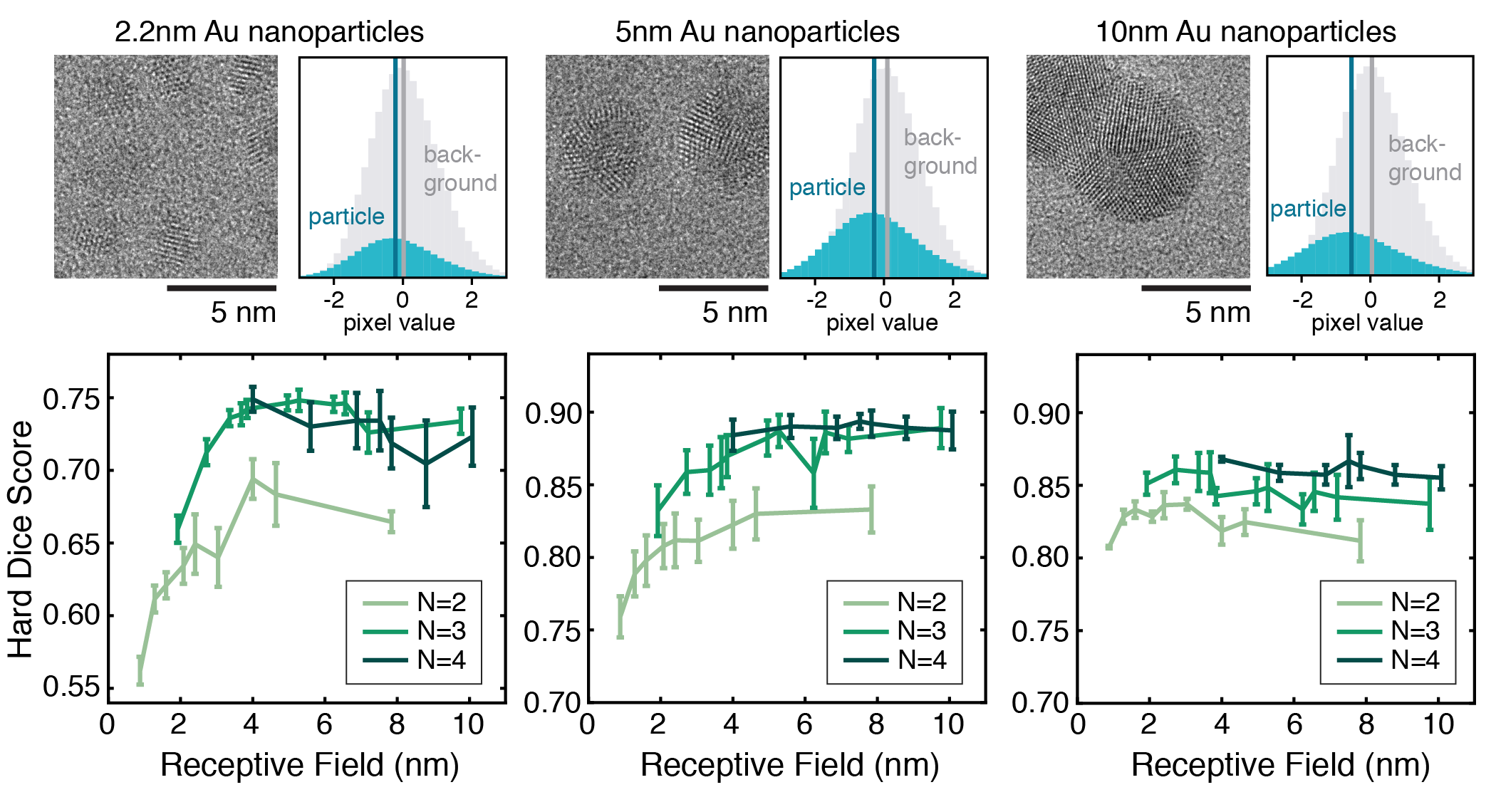}
    \caption{Receptive field dependence in high-resolution TEM datasets of 2.2nm, 5nm, and 10nm Au nanoparticles. For each dataset, we show a sample image from the test set, the pixel value histograms of the training set separated by label, and the segmentation performance as a function of receptive field and network complexity. }
    \label{fig:hrtemRF}
\end{figure}

This receptive field dependence is seen in all high-resolution TEM images, regardless of nanoparticle size. We repeat the receptive field experiment on three new high-resolution TEM datasets taken at the same magnification, each of either 2.2nm, 5nm, or 10nm Au nanoparticles on an ultrathin carbon substrate. In all three datasets, segmentation performance increases then plateaus with larger receptive field, though the dependence becomes less noticeable as nanoparticle size increases (Figure \ref{fig:hrtemRF}). Interestingly, the receptive field value at which performance starts to plateau is greatest for the small (2.2nm) nanoparticles, requiring a receptive field much greater than the average diameter of the nanoparticles for peak performance. We hypothesize that the inverse relationship between nanoparticle size and necessary receptive field is due to the greater nanoparticle contrast as diameter (and therefore, thickness) increases. By comparing the histograms of pixel values in the three datasets, we see that the contrast between nanoparticle and background increases with larger nanoparticle size. As receptive field is less important in high contrast images, as noted in Figure \ref{fig2}c, it is likely that as the nanoparticles become easier to identify, the network requires less spatial information to gauge changes in contrast, and therefore there is less dependence on receptive field.

\begin{figure}[h]
    \centering
    \includegraphics{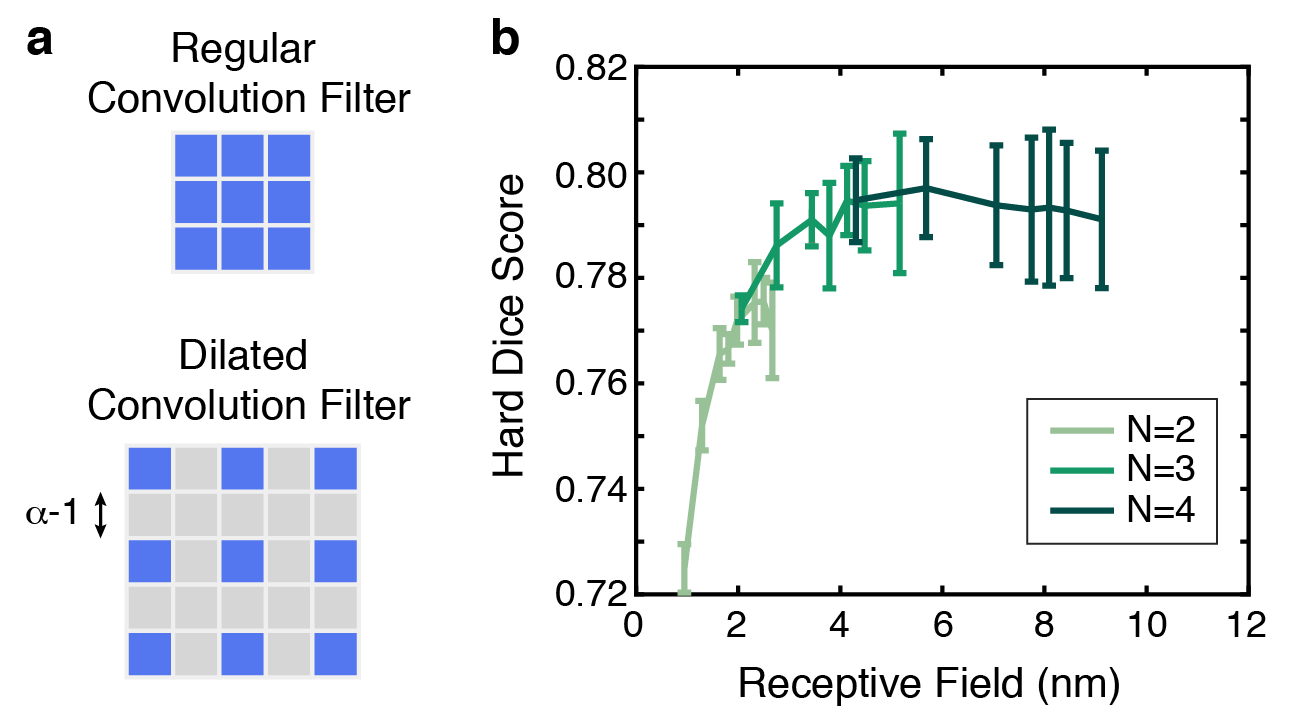}
    \caption{Neural network performance when increasing the receptive field using dilated convolutions. (a) Schematic depicting how dilation changes the convolution filters. (b) Segmentation performance on high-resolution TEM images of 5nm Au nanoparticles when receptive field is increased using dilated convolutions. }
    \label{fig:dilation}
\end{figure}

In addition to modifying the max pooling layers, there are multiple strategies that can modify receptive field, including changing hyperparameters like filter size, stride, and/or number of layers. To further demonstrate that our results are not limited to this max pooling strategy, we repeat our receptive field investigation on the high resolution TEM images of 5nm Au nanoparticles, but now vary the dilation parameter of our convolutional filters (Figure \ref{fig:dilation}a) instead of the max pooling kernel size. As seen in Figure \ref{fig:dilation}b, as receptive field is increased using dilated convolution, we observe an increase in segmentation performance, followed by a plateau around 5nm, similar to our results on the same dataset using max pooling (Figure \ref{fig2}d). Quantitatively, both methods similarly saturate around 0.79 in hard dice score. The similarities in behavior despite technical differences in the neural network architecture further cements that receptive field is the key parameter affecting performance.

\section*{Discussion}

In contrast-based low-resolution TEM images, the receptive field is not an important factor in segmentation performance. The network does not show any increase in performance as it changes from a receptive field smaller than half of the nanoparticle diameter to a much larger receptive field size. These results suggest that our networks are not solely relying on edge-detection. An network that relies on edge-detection would likely show an increase in performance when the receptive field is about half of the nanoparticle diameter since smaller receptive fields would not be able to detect the nanoparticle edges at all points in a nanoparticle. 

\begin{figure}[h]
    \centering
    \includegraphics[width=\textwidth]{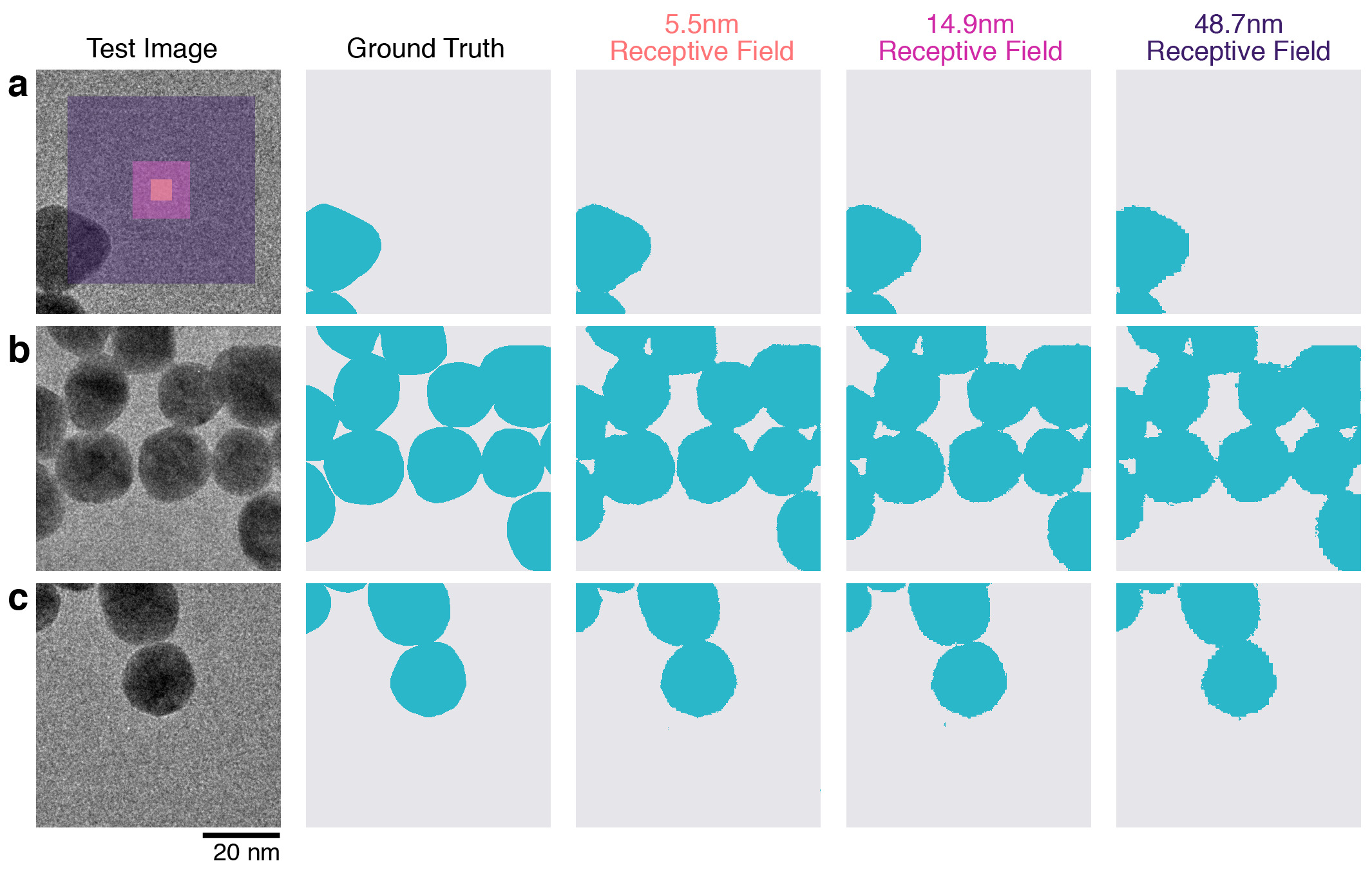}
    \caption{Example segmentation results on low-resolution TEM images of 20nm Au nanoparticles for a 2-residual-block network with different receptive fields. Colored blocks superimposed on the TEM image in (a) show the relative size of the receptive fields of the three networks.}
    \label{fig:titanEx}
\end{figure}

The slight decrease in performance with increasing receptive field can be attributed to aliasing effects from the larger max pooling filters. Since max pooling takes the maximal value in a $k \times k$ pixel area (and its corresponding upsampling procedure repeats the maximal value in a $k \times k$ area), we lose fine detail information as $k$ increases. This can be qualitatively seen in Figure \ref{fig:titanEx}, which shows how three 2-residual-block networks, which only differ by their receptive fields, perform when segmenting three test images. We see that qualitatively, all three networks correctly segment the nanoparticles, but the results from the 48.7nm receptive field network have rough, blockier edges, which we attribute to the large max pooling filter size. Note that in practice, max pooling is often kept to $k=2$ to avoid these blocky artifacts. 

\begin{figure}
    \centering
    \includegraphics[width=\textwidth]{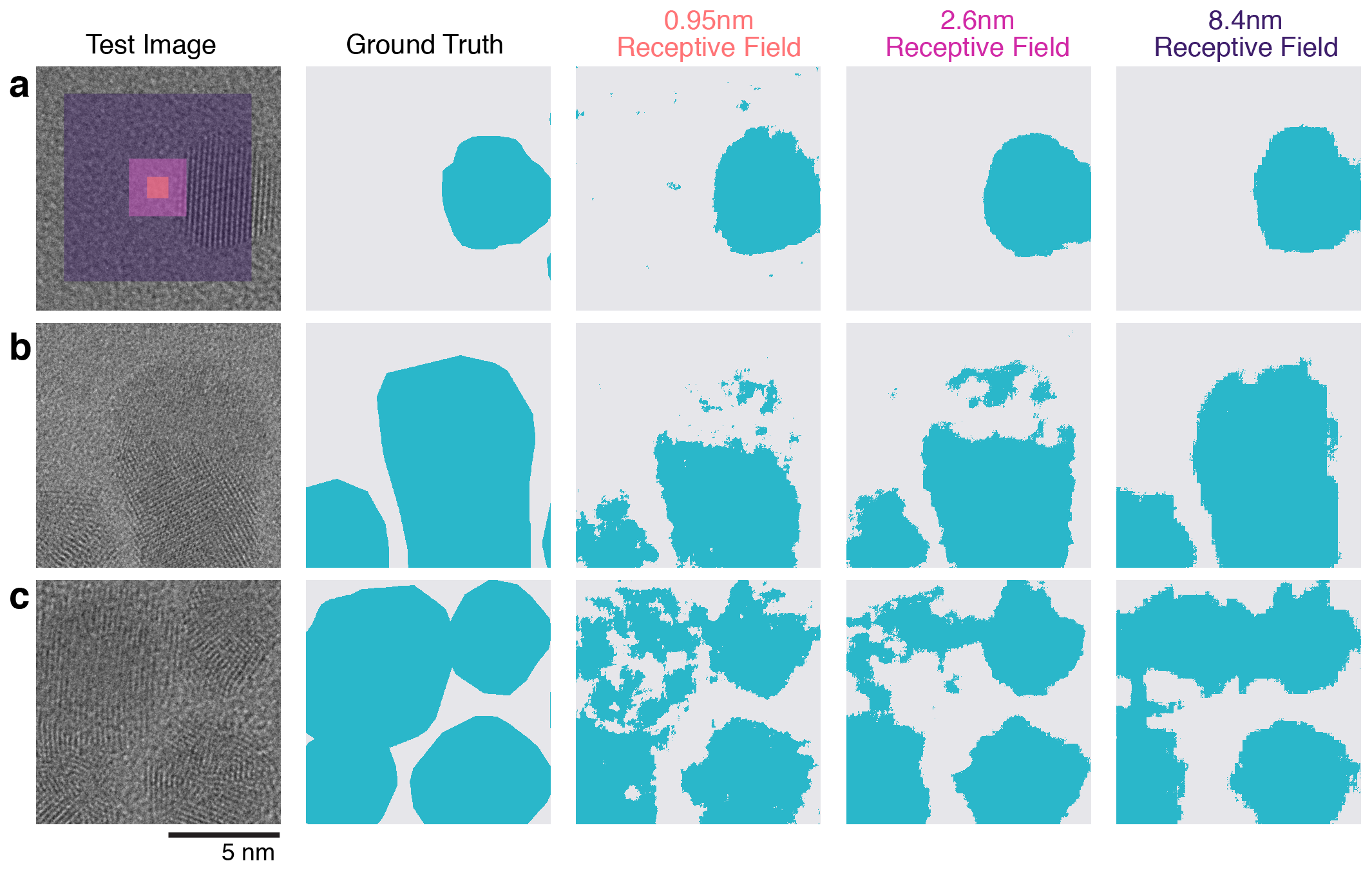}
    \caption{Example segmentation results on high-resolution TEM images of 5nm Au nanoparticles for a 2-residual-block network with different receptive fields. Colored blocks superimposed on the TEM image in (a) show the relative size of the receptive fields of the three networks. }
    \label{fig:AuKateEx}
\end{figure}

High-resolution TEM images, however, quantitatively and qualitatively show a significant difference as receptive field is increased. In Figure \ref{fig:AuKateEx}, we again compare segmentation results from three 2-residual-block networks of various receptive fields. For an ideal high-resolution TEM image in which the lattice fringes are visible for the entire nanoparticle (Figure \ref{fig:AuKateEx}a), all three networks perform equally well. The network with the smallest receptive field occasionally misclassifies parts of the background region as nanoparticle but the larger receptive field networks do not make the same mistake. The results from the 8.4nm receptive field network also show the same blocky artifacts seen in the 48.7nm receptive field network in Figure \ref{fig:titanEx}; since these two networks have the exact same architecture, we further confirm that these artifacts are from the max pooling filter size. 

The small receptive field network also misclassifies nanoparticle regions where there are fainter or no visible lattice fringes, but larger receptive field networks are able to segment those same regions correctly (Figure \ref{fig:AuKateEx}b,c). By human eye, these regions are still identified as part of the nanoparticle due to both the slight change in contrast from the background and contextual information about the nanoparticle shape (i.e. spherical and convex). This again supports our findings in Figure \ref{fig:hrtemRF} that larger receptive fields enable better segmentation of low-contrast nanoparticles. 

\begin{figure}
    \centering
    \includegraphics[width=\textwidth]{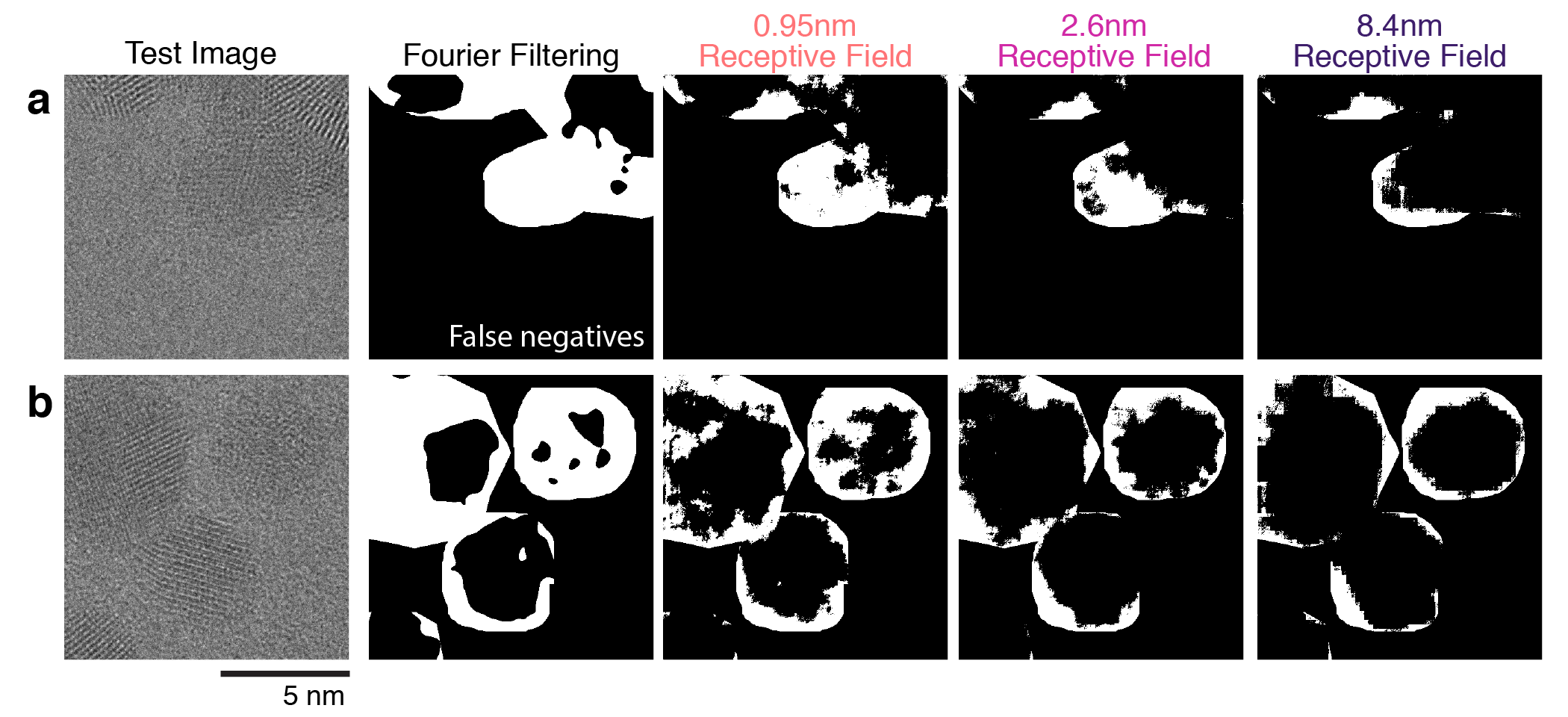}
    \caption{Comparing false negative regions in UNet segmentation results against false negative regions from Fourier filtering. (a,b) Example test images and their corresponding false negative maps after segmentation via Fourier filtering and three 2-residual-block networks with varying receptive fields. }
    \label{fig:ffex}
\end{figure}

These misclassified nanoparticle regions, denoted as false negatives, happen to be areas that Fourier filtering, a purely texture-only image segmentation technique for high-resolution TEM images, also miss. To compare, we manually Fourier filter the non-augmented test images and compare their false negative regions against the false negative regions from the 0.95nm, 2.6nm, and 8.4nm receptive field networks (Figure \ref{fig:ffex}a,b). All segmentation techniques struggle with nanoparticle edges, but the false negative regions from Fourier filtering look more similar to the 0.95nm receptive field results than the 8.4nm receptive field results. For most of the test images, the smaller receptive field network has false negatives most similar to Fourier filtering (see Figure S7 for statistics), suggesting that smaller receptive field networks learn a non-linear function similar to Fourier filtering. We hypothesize that the spatial constraints force the neural network to learn the simplest way to segment with limited contextual clues - by identifying lattice fringes. Once that spatial constraint is expanded, the network learns to incorporate macroscopic contrast information and therefore improves performance. 

If small receptive fields lead to trained neural networks with results similar to Fourier filtering, then we also have an alternative explanation as to why small receptive field networks still perform well on large (10nm) Au nanoparticles. When using Fourier filtering to segment, the 10nm nanoparticle dataset has the highest dice score (Table S3), likely because it has more areas with visible lattice fringes. Therefore, it does not benefit as much from modifying the neural network's learned function from something similar to Fourier filtering to something more complex. 

Similar to prior denoising results, we also find that the receptive field needs to be increased to deal with the larger (in pixel size) image features in high-resolution TEM images \citep{vincent2021developing}. Given the differences in image task (denoising versus segmentation) and network architecture but similar results with respect to receptive field, we conclude that receptive field is an important consideration when working with high-resolution TEM images. The plateauing behavior with increased receptive field suggests that receptive field is the key parameter that limits performance in lightweight networks, and so to increase neural network performance, the receptive field should also be considered in addition to network complexity. This finding provides a strategy to increase performance of lightweight networks without requiring more TEM data or new architectures, broadening the number of usable networks. 

\begin{figure}[h]
    \centering
    \includegraphics{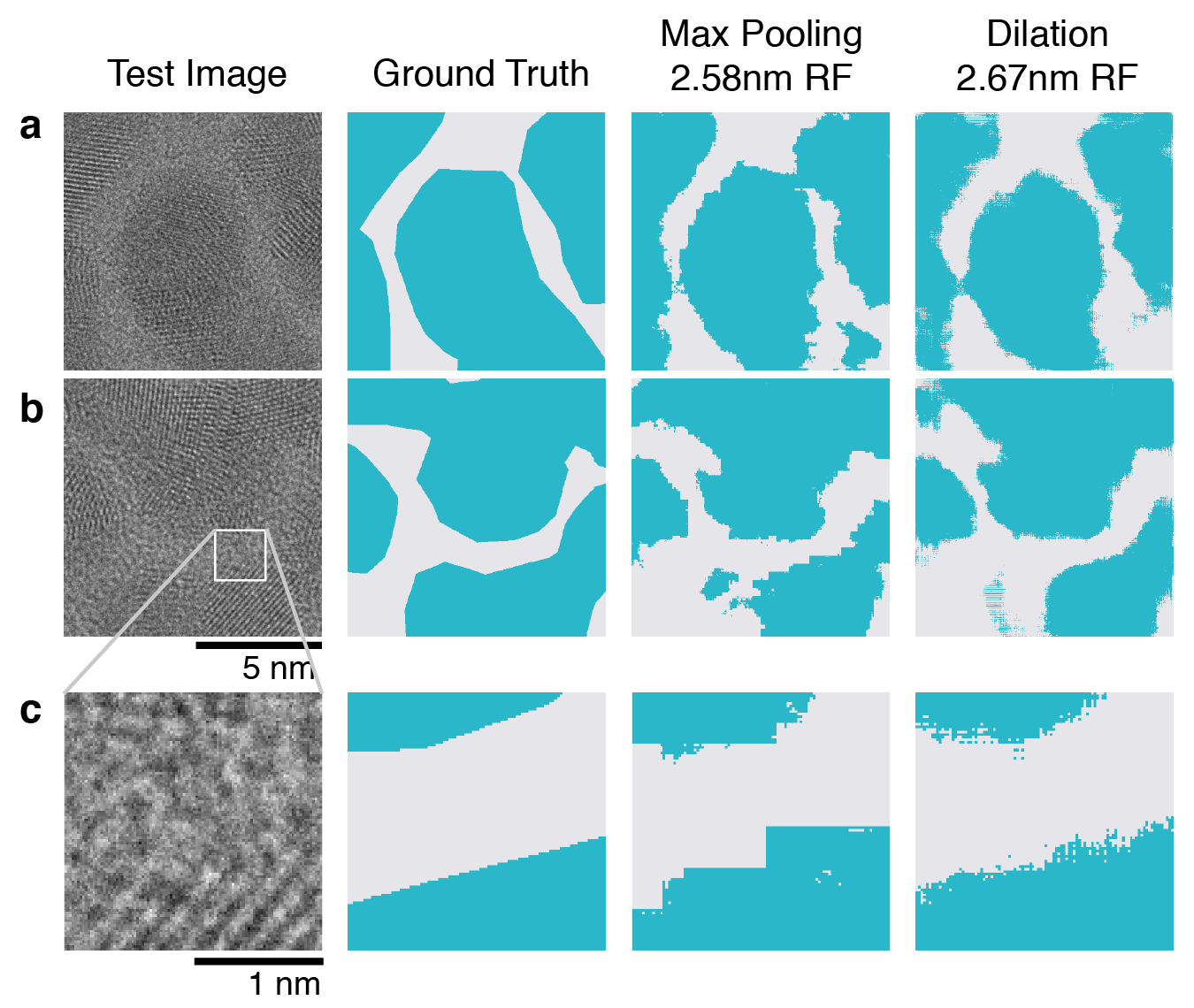}
    \caption{Comparison of segmentation results for a $\sim$ 2.6nm receptive field using either max pooling or dilation. (a,b) Two test images from the high-resolution dataset of 5nm Au nanoparticles and their respective segmentation results. (c) A zoom-in of the test image in (b) to highlight the edge artifacts from both max pooling and dilation. }
    \label{fig:dilation_examples}
\end{figure}

In practice, the choice of how to increase receptive field will depend on the image analysis task and the acceptable types of image artifacts. By qualitatively comparing performance between a neural network with increased receptive field using max pooling versus dilated convolution, we see that that these different strategies lead to similar segmentation results but different image artifacts at the nanoparticle edges (Figure \ref{fig:dilation_examples}). Both networks misclassify similar regions, as seen in the bottom right of the test image in Figure \ref{fig:dilation_examples}a and bottom left of the test image in Figure \ref{fig:dilation_examples}b, suggesting that the two networks are identifying similar features. If we zoom in (Figure \ref{fig:dilation_examples}c), the larger max pooling kernel size leads to blocky edges, as noted before, while dilation leads to gridding artifacts near the nanoparticle edges which make it difficult to identify the exact location of the edge. Therefore, for a dense image analysis task like nanoparticle segmentation where information about the edges are important, one would likely avoid using max pooling and dilation to increase receptive field, but instead utilize larger convolution filters, more layers, or other novel ideas being proposed in computer vision \citep{wang2021smoothed}. On the other hand, for a classification task which associates an entire image with a label, the choice in strategy may not matter as much.

Finally, we note that the receptive field reported here is the theoretical \textit{maximal} area that contributes to the final decision. In reality, the receptive field is not equally weighted and the effective receptive field, or the area that significantly influences the decision, is smaller \citep{luo2016understanding}. The effective receptive fields of our trained UNets are dominated by pixels nearby the decision pixel due to skip connections, even as the maximal receptive field is increased (Figure S8). Further examination shows that the edges of the larger receptive field networks still contribute, and we hypothesize that these large receptive field networks maintain their high performance because they incorporate both local and global information. 

\section*{Summary}
In summary, by systematically modifying the receptive field for various combinations of neural network complexities and TEM image datasets, we have identified how neural network constraints affect nanoparticle segmentation. While low-resolution, contrast-based TEM images are insensitive to the size of a neural network's receptive field, high-resolution contrast- and texture-based TEM images require neural networks with a large enough receptive field in order to perform well. Receptive field is especially important when segmenting small and/or low-contrast nanoparticle regions, as only large-enough receptive fields can detect the subtle change in contrast. Our results provide intuition as to how neural network architecture choices affect TEM image analysis and guidance for microscopists interested in customizing neural network architectures for their datasets. 

\section*{Data Availability}
All datasets and ground truth labels acquired for this paper are available on Zenodo: \href{https://doi.org/10.5281/zenodo.6419024}{https://doi.org/10.5281/zenodo.6419024}. Code and Jupyter notebooks are available at \href{https://github.com/ScottLabUCB/HRTEM-Receptive-Field}{https://github.com/ScottLabUCB/HRTEM-Receptive-Field}.

\section*{Acknowledgements}
This work was primarily funded by the US Department of Energy in the program “4D Camera Distillery: From Massive Electron Microscopy Scattering Data to Useful Information with AI/ML". Imaging was done at the Molecular Foundry, which is supported by the Office of Science, Office of Basic Energy Sciences, of the U.S. Department of Energy under Contract No. DE-AC02-05CH11231. Part of this research used the Savio computational cluster provided by the Berkeley Research Computing program at the University of California, Berkeley (supported by the UC Berkeley Chancellor, Vice Chancellor for Research, and Chief Information Officer).

\section*{Competing Interests}
The authors declare no competing interests.

%\section*{References}
\bibliographystyle{MandM}
%\begin{thebibliography}{}
%\bibitem{aguiar2019decoding}
%Aguiar, J., Gong, M. L., Unocic, R., Tasdizen, T., and Miller, B., “Decoding crystallography from high-resolution electron imaging and diffraction datasets with deep learning,” Science Advances 5, eaaw1949 (2019).
%\end{thebibliography}

\bibliography{refs}    

\end{document}